\documentstyle[preprint,pra,aps]{revtex}
\tightenlines
\begin{document}
\draft
\title{\bf Calculation of parity and time invariance violation
in the radium atom.}
\author{V.A.Dzuba, V.V.Flambaum and J.S.M. Ginges}
\address{School of Physics, University of New South Wales, 
Sydney 2052,Australia}
\date{\today}
\maketitle

\begin{abstract}
Parity ($P$) and time ($T$) invariance violating effects in the Ra atom 
are strongly enhanced due to close states of opposite parity, the large 
nuclear charge Z and the collective nature of $P,T$-odd nuclear moments. 
We have performed calculations of the atomic electric dipole moments (EDM) 
produced by the electron EDM and the nuclear magnetic quadrupole and Schiff 
moments. We have also calculated the effects of parity non-conservation 
produced by the nuclear anapole moment and the weak charge. Our results 
show that as a rule the values of these effects are much larger than those 
considered so far in other atoms (enhancement is up to $10^5$ times).

\end{abstract} 
\vspace{1cm}
%***************************************************************************

\section{Introduction}

The lower energy levels of radium corresponding to configurations of 
different parity have very close energies. This leads to a strong 
enhancement of the various parity ($P$) and time ($T$) invariance violating 
effects.
In our previous paper \cite{Flambaum99}
we considered the states $7s6d~^3\mbox{D}_2$ with $E=13993.97 \mbox{cm}^{-1}$ 
and $7s7p~^3\mbox{P}_1$ with $E=13999.38 \mbox{cm}^{-1}$, which are separated 
by a very small interval of $\sim$5 cm$^{-1}$ ($\sim 10^{-3}$ eV). 
Simple estimates showed that 
the effects of nuclear $P$- and $T$-odd moments such as the magnetic 
quadrupole moment (MQM), the Schiff moment (SM) and the anapole moment (AM) 
are many times larger than in all atomic systems considered before.
In the present paper we present more accurate calculations of these and  
other parity and time invariance non-conserving effects in those states of 
the radium atom where the effects are large. 
We use a relatively simple {\it ab initio} approximation to perform the 
calculations. The approximation is a reasonable compromise between the 
simplicity of the calculations and the accuracy of the results.
It is based on relativistic Hartree-Fock (RHF) and configuration 
interaction (CI) methods. A minimum number of basis states are used at the 
CI stage of the calculations. However, important many-body effects, such as 
polarization of the atomic core by an external field and
correlations between core and valence electrons, are included in the 
calculations of single-electron matrix elements.
To control the accuracy of the calculations we also 
calculated hyperfine structure intervals and lifetimes of lower states
of radium and its lighter analog barium. 

Our calculations confirm the estimates done in the previous work 
\cite{Flambaum99} and show that the value of most $P$- and $T$-odd effects in
radium is much higher than in other atoms considered before. 
The parity non-conserving (PNC) electric dipole transition amplitude between 
the ground and $^3$D$_1$ even states is about
$E1_{PNC} \approx 0.8 \times 10^{-9} (Q_W/N) iea_0,$
which is 100 times larger than the measured PNC amplitude in cesium 
\cite{Wood} and about 5 times larger than the corresponding amplitude 
in francium \cite{Dzuba95}.
The enhancement of the electron electric dipole moment (EDM) in the 
$^3$D$_1$ state of Ra is about 5400, which is again many times larger
than corresponding values for the ground states of Fr (910) and Au (260)
\cite{Byrnes}.
The transition amplitude between the ground and $^3$D$_2$ even states 
induced by the nuclear anapole moment is about $10^{-9} e a_0$, which is 
more than $10^3$ times larger than a similar amplitude in Cs \cite{Wood}.
Also, the EDM of the Ra atom in the $^3$D$_2$ state induced by the nuclear
Schiff and magnetic quadrupole moments is strongly enhanced. Both
contributions (SM and MQM) are about $10^{-19} \eta~e\cdot cm$
($\eta$ is the dimensionless constant of the $P$-,$T$-odd nucleon-nucleon
interaction).
This is again about $10^5$ times larger than the EDM of the Hg atom
which currently gives the best limit on $\eta$ \cite{Jacobs}.
All this makes radium a very promising candidate for the experimental 
study of $P$- and $T$-odd forces by means of atomic physics.

\section{Method} 
\label{theory}

We use relativistic Hartree-Fock (RHF) and configuration interaction (CI)
methods to construct two-electron wave functions of the ground and lower
excited states of barium and radium. The calculations start from the RHF 
method for a closed shell system corresponding to the ground state
configuration ($6s^2$ for Ba and $7s^2$ for Ra). 
Since $nsnp$ and $ns(n-1)d$ configurations, with $n=6$ for Ba and $n=7$ 
for Ra, do not correspond to a closed-shell system, we calculate $p$ and 
$d$ basis states in a model HF potential. For example, to calculate $7p$ 
and $6d$ states of Ra,
we keep all other states frozen, remove the contribution of one $7s$-electron 
from the direct HF potential and use this potential to calculate the 
required states. The same procedure applies for Ba.
Thus, we have five single-electron basis states for the
CI calculations ($ns_{1/2}, np_{1/2}, np_{3/2}, (n-1)d_{3/2}, (n-1)d_{5/2}$). 
It turns out, however, that this simple CI approximation significantly 
overestimates the relative value of spin-orbit intervals for the odd-parity 
states and underestimates it for the even-parity states. This affects the 
accuracy of the calculation of $P$- and $T$-odd effects because most of them 
involve transitions with a change of spin which are sensitive to the value 
of the relativistic effects. We found that the spin-orbit intervals are 
sensitive to the screening
of the Coulomb interaction between two external electrons
(recall that Coulomb integrals contribute to the spin-orbit splitting
due to the difference between the single-particle radial
wave functions belonging to different components of the single-particle 
doublets).
To improve the quality of the wave functions we introduce fitting
factors $f_k$ to the Coulomb interaction in the CI calculations
($k$ is the multipolarity of the Coulomb interaction). It was found that 
multiplying all Coulomb integrals of multipolarities 0, 1 and 2 by
factors $f_0 = 0.7, f_1 = 0.75, f_2 = 0.9$ significantly improves the 
energies and fine structure intervals of lower odd and even states of 
barium and radium. These factors simulate the effect of the screening
of the Coulomb interaction between valence electrons and core electrons.
They also compensate to some extent the effect of the incompleteness of the
basis set.

To calculate values other than energy, such as the effect of electron
interaction with photons and nuclear $P$- and $T$-odd fields, we also
 include core polarization effects (direct and exchange RPA-type corrections) 
and core-valence correlation effects (the Bruckner-type correlation 
corrections). These two effects are very important for the considered states 
of radium. 
Indeed, consider mixture of the $^3\mbox{D}_J$ and $^3\mbox{P}_{J'}$
states by the $P$- and $T$-odd interaction $W$. 
Corresponding dominant configurations ($7s6d$ and $7s7p$) can 
only be mixed by a $\langle 7p|W|6d \rangle$ matrix element.
However, this matrix element is extremely 
small in the Hartree-Fock approximation. This is because the electron 
interaction with $P$- and $T$-odd nuclear moments is localized in the 
vicinity of the nucleus
where the $d$-electron does not penetrate due to the centrifugal barrier.
On the other hand, the polarization of the electron core by these moments
 produces a long-range correction $\delta V$ to the HF potential which 
effectively renormalizes the interaction of an external electron with the 
nucleus. The corresponding matrix element $\langle 7p|W + \delta V |6d 
\rangle$ is not small even in the case of the $p-d$ transition due to 
the long-range of the renormalized interaction $W + \delta V$. Note that 
$\langle 7s|W|7p \rangle$ matrix elements
also contribute to the mixture of the $^3\mbox{D}_J$ and $^3\mbox{P}_{J'}$
states due to the configuration interaction. 
Thus, there is an interference of several factors: the $s-p$ matrix 
elements are large but their contribution is suppressed due to the 
smallness of the configuration mixing. The $p-d$ matrix elements are 
considerably smaller (although not negligible) but they appear in the 
dominating configurations.
It cannot be said in advance which transitions are more important and
as we see from our calculations there are cases when $s-p$ transitions
dominate over $p-d$ and vice-versa (see below).
The Bruckner-type correlation corrections (the correlation corrections to 
the single-electron wave functions) are also important, since they
increase the density of an external electron on the nucleus by 
$\sim 30\%$ (see e.g. \cite{Dzuba84}).

The full scale inclusion of the core polarization and correlation
 effects into the CI calculations (see, e.g. \cite{Kozlov}) lies beyond the 
framework of this research. We adopted a simplified approach in
which the corresponding corrections are calculated for the single-particle
matrix elements. The relative values of the renormalization of the matrix
elements by the core polarization and core-valence correlations have been 
extrapolated from accurate calculations of the core polarization and 
Bruckner-type correlation corrections for the radium positive ion. 
Ra$^+$ has a simple electronic structure - one electron above closed 
shells - and the corresponding procedures are well defined for it 
\cite{Dzuba83}.

To check our method and the accuracy of our results, we calculated the 
hyperfine structure (hfs) constants of $^{213}$Ra and $^{137}$Ba. 
The results for the energies and hfs constants are presented in 
Table \ref{BaRa}. 
One can see that even for this very simple CI approximation the accuracy 
of the energies and fine structure intervals is very good. 
The accuracy of the hfs constants is also good for the most important
states $^3$D$_2$ and $^3$P$_1$.
Table \ref{rad} shows the effect of the core polarization (RPA) and
Bruckner-type correlations ($\Sigma$) on the single-electron matrix 
elements. One can see that these effects play a crucial role in the
$p-d$ matrix elements. However, their contribution to the $s-p$
matrix elements is also very important.

\section{Parity Violation in $7s^2 \rightarrow 7s6d$ transitions}

\subsection{Spin-independent parity non-conservation}

The Hamiltonian $H_{PNC}$ of the interaction of an electron with the nuclear 
weak charge $Q_W$ (formula (\ref{APNC}) in the appendix)
mixes states of the same total momentum $J$ and opposite
parity. Thus, electric dipole transitions between states of initially
equal parity become possible. In particular, the transition between the 
ground state
$^1$S$_0$ and the excited $^3$D$_1$ state is enhanced due to the closeness
of the opposite parity state $^1$P$_1$.
The dominating contribution to this transition is given by
\begin{equation}
E1_{PNC} = \frac{\langle 7s^2~^1\mbox{S}_0 |d_z| 7s7p~^3\mbox{P}_1 \rangle
\langle 7s7p~^3\mbox{P}_1|H_{PNC}| 7s6d~^3\mbox{D}_1 \rangle}
{E(^3\mbox{D}_1) - E(^3\mbox{P}_1)}.
\label{E1PNC}
\end{equation}
Apart from the enhancement, there are several suppression factors in 
(\ref{E1PNC}).
First, the electric dipole matrix element is small because of a change of spin.
It is 3 to 5 times smaller than most of those amplitudes which do not
change atomic spin (see Table \ref{E1}).
Second, in the matrix element of the PNC interaction, leading 
configurations produce only the $p_{3/2} - d_{3/2}$ single-electron matrix 
element which is small. It is not zero mostly due to core polarization. 
However, it is about 25 times smaller than the $s_{1/2} - p_{1/2}$ matrix 
element.
The latter contribute to the PNC amplitude due to configuration mixing.
Our calculations show that the contribution of the $s - p$ transition 
to the PNC amplitude is about 7 times larger than the contribution of the 
$p - d$ transition.
In spite of some suppression, the final answer is quite large:
\begin{eqnarray}
 ^{225}\mbox{Ra}: E1_{PNC}& =& 0.77 \times 10^{-9} (Q_W/N) iea_0,\\
 ^{223}\mbox{Ra}: E1_{PNC}& =& 0.76 \times 10^{-9} (Q_W/N) iea_0.
\label{RPNC}
\end{eqnarray}
This is one hundred times larger than the measured PNC amplitude in cesium 
\cite{Wood} and about 5 times larger than the corresponding amplitude 
in francium \cite{Dzuba95}.
Even radium isotopes have close values of the amplitudes (approximately,
the effect is proportional to the number of neutrons $N$).

\subsection{Anapole moment}

The Hamiltonian of the electron interaction with the nuclear anapole 
moment is presented in the appendix (\ref{AAM}).
Similar to the spin-independent PNC interaction, it
mixes states of opposite parity and leads to non-zero E1-transition amplitudes
between states of initially equal parity. However, it can also mix states with
$\Delta J =1$ and it depends on the nuclear spin, so that its contribution to
transitions between different hyperfine structure components are
different. The corresponding expression is very similar to (\ref{E1PNC}).
However, dependence on the hyperfine structure must be included 
(see formula (\ref{AE1AM}) in the appendix for details). This amplitude is 
proportional to the
$\langle ^3\mbox{P}_1 ||\vec \alpha \rho(r)||^3\mbox{D}_2 \rangle$ 
matrix element. 
Contributions of different single-electron transitions into this
matrix element 
are presented in Table \ref{HME}. Note the strong cancellation between terms 
corresponding to $s - p$ and $p - d$ transitions. This means that an
accurate inclusion of the core polarization and core-valence correlation
effects is very important indeed, as has been discussed above. We believe 
that the fitting of the energies helps to stabilize this matrix element 
similar to the case of the $E1$-transition amplitude.

The results for $^1$S$_0 - ^3$D$_1$ and $^1$S$_0 - ^3$D$_2$ transitions
are presented in Table \ref{AnM}. Note that the contribution of the anapole
moment to the PNC amplitude (\ref{RPNC}) can be measured by comparing the
amplitudes between different hyperfine structure components similar to
what was done for cesium \cite{Wood}. However, it may be
much more efficient
to measure the effect of the anapole moment in the $^1$S$_0 - ^3$D$_2$ 
transition
because it is about ten times larger due to the small energy denominator and
because the nuclear spin independent PNC interaction does not contribute
to this amplitude at all due to the large change of the total electron
angular momentum
$\Delta J =2$.

\section{Atomic Electric Dipole Moments}

\subsection{Electron EDM}

An electron electric dipole moment interacting with an atomic field mixes
states with the same total momentum $J$ and opposite parity. As a result,
an atomic EDM appears. The EDM of radium in the  
$^3\mbox{D}_1$ state is strongly enhanced due to the closeness of the opposite
parity state $^3\mbox{P}_1$.
In an approximation when only the mixture of the closest states is included, 
the EDM is given by
\begin{eqnarray}
	d=2\frac{\langle7s6d~^3\mbox{D}_1|-e{\mathbf r}|7s7p~^3\mbox{P}_1 
	\rangle  \langle 7s7p~^3\mbox{P}_1|H_{EDM}|7s6d~ ^3\mbox{D}_1 \rangle}
	{E(^3\mbox{D}_1)-E(^3\mbox{P}_1)}.
\label{FEDM}
\end{eqnarray}
Calculations using formulae from the appendix give the following result
\begin{eqnarray}
	&  d=5370 d_e.
\label{REDM}
\end{eqnarray}
Note that a very strong enhancement is caused by the small
energy denominator $E(^3\mbox{D}_1)-E(^3\mbox{P}_1)$ = 0.001292 a.u.
%Note also that about the same enhancement of the electron EDM takes place
%in the $^3$P$_1$ state of radium.

\subsection{Schiff moment}

Electron interaction with the nuclear Schiff moment also produces
an atomic EDM. The EDM of Ra caused by Schiff moment is strongly enhanced
in the $^3$D$_2$ state. Its value is approximately given by
\begin{equation}
	d_z = 2 \frac{\langle 7s6d~^3\mbox{D}_2|d_z|7s7p~^3\mbox{P}_1 \rangle
	\langle 7s7p~^3\mbox{P}_1|H_{SM}| 7s6d~^3\mbox{D}_2 \rangle}
	{E(^3\mbox{D}_2) - E(^3\mbox{P}_1)}.
\label{EDMSM}
\end{equation}
More detailed formula which include the dependence of (\ref{EDMSM}) on the 
hyperfine structure is presented in the appendix (\ref{AEDMSM}).

Table \ref{HME} shows single-electron contributions to the
$\langle ^3\mbox{P}_1 ||H_{SM}||^3\mbox{D}_2 \rangle$ 
matrix element. Note that $s - p$ transitions strongly dominate here.
However, the contribution of the $p - d$ transitions is not negligible and
should be included for accurate results.

Calculated values of the radium EDM induced by the Schiff moment are presented
in Table \ref{RSM}.

\subsection{Magnetic quadrupole moment}

Electron interaction with the nuclear MQM can also produce an EDM of an atom. 
However, in contrast with the case of the Schiff moment, the MQM of isotopes 
where the nuclear spin $I<1$ (like $^{225}$Ra, where $I =1/2$) is zero.
The EDM of Ra in the  $^3\mbox{D}_2$ state is given by a formula similar
to (\ref{EDMSM})
\begin{equation}
	d_z = 2 \frac{\langle 7s6d~^3\mbox{D}_2|d_z|7s7p~^3\mbox{P}_1 \rangle
	\langle 7s7p~^3\mbox{P}_1|H_{MQM}| 7s6d~^3\mbox{D}_2 \rangle}
	{E(^3\mbox{D}_2) - E(^3\mbox{P}_1)}.
\label{EDMMQM}
\end{equation}
Again, more detailed formula can be found in the appendix (\ref{AEDMMQM}).

Table \ref{HME} shows single-electron contributions to the
$\langle ^3\mbox{P}_1 ||A_{mk}||^3\mbox{D}_2 \rangle$ 
matrix element. Note that in contrast to the cases of the Schiff and
anapole moments, $p - d$ transitions dominate over $s - p$ transitions in 
this matrix element.
For the anapole moment these two types of transitions contribute almost 
equally, while for the Schiff moment $s -p$ transitions dominate.
Note that $s - p$ transitions appear due to configuration mixing only,
while contribution of the $p - d$ transitions is extremely small if core
polarization is not included.
This indicates once more that even for a rough estimation of the time or
parity invariance violating effects in Ra an inclusion of the appropriate
many-body effects is essential.

Calculated values of the radium EDM induced by the magnetic quadrupole 
moment are presented in Table \ref{RMQM}.

\section{Lifetimes}

To plan experimental measurements of space and time invariance violation 
in radium it is important to know the lifetimes of the states of interest. 
Apart from that, comparison of the calculated and experimental lifetimes
can serve as a good test of the method used for calculation of $P$- and $T$-
invariance violation since the same dipole transition amplitudes contribute 
in either case. 
As far as we know, none of the radium lifetimes have been measured so far. 
On the other hand, some experimental data is available for barium.
Therefore, we calculated lifetimes of lower states of both atoms.
Results for dipole transition amplitudes are presented in Table \ref{E1}
and the corresponding lifetimes are in Table \ref{tau}.

For the purpose of the present work, the most important states of radium 
are $^3$P$_1$,$^3$D$_1$ and $^3$D$_2$ states. The decay rate of the $^3$P$_1$
state is strongly dominated by the $^3$P$_1$ - $^1$S$_0$ transition.
Transitions to the $^3$D$_1$ and $^3$D$_2$ states are suppressed due to
small frequencies. The $^3$P$_1$ - $^1$S$_0$ dipole transition amplitude
involves a change of the atomic spin and therefore is sensitive to the
value of the relativistic effects. This makes the amplitude numerically 
unstable. This probably explains the poor agreement between different 
calculations (see Table \ref{tau}). However, we believe that 
the fitting of the fine structure which we have done for Ba and Ra
(see Section \ref{theory}) brings the amplitude close to the correct value.
This is supported by similar calculations for barium. The
$^3$P$_1$ - $^1$S$_0$ amplitude contributes 38\% to the decay rate of
the $^3$P$_1$ state of barium. Good agreement between calculated and
experimental lifetimes of this state (see Table \ref{tau}) means that
all transition amplitudes, including the $^3$P$_1$ - $^1$S$_0$ amplitude,
are calculated quite accurately.

The lifetime of the $^3$D$_1$ state of Ra is determined by the $^3$D$_1$ -
$^3$P$_0$ transition which is numerically stable. The lifetime of this 
state calculated by us is in good agreement with the estimations 
done by Budker and DeMille \cite{Budker}. 

The $^3$D$_2$ state of radium
is a metastable state. It decays only via electric quadrupole (E2)
transition to the ground state. Calculations similar to the electric
dipole transitions show that the lifetime of this state in the absence 
of external fields is about 15 seconds. However, measurements of the
atomic EDM involve placing the atoms in a strong electric field.
It is important to know how the lifetimes of the $^3$D$_2$ and
$^3$D$_1$ states of Ra are affected by this field.
The electric field mixes states of different parity and $\Delta J = 0,\pm 1$. 
If only an admixture of the nearest state is taken into account, 
the amplitude which determines the decay rate of a $^3$D$_J$ state is
given by
\begin{equation}
	A = \frac{\langle ^1\mbox{S}_0|d_z {\mathcal E}|^3\mbox{P}_1 \rangle 
                  \langle ^3\mbox{P}_1 | d_z | ^3\mbox{D}_{J} \rangle}
	{E(^3\mbox{P}_1)-E(^3\mbox{D}_{J})}.
\label{Atau}
\end{equation}
Where ${\mathcal E}$ is the electric field.
This leads to the following decay rates
\begin{eqnarray}
 && W(^3\mbox{D}_2) = 0.21 {\mathcal E}^2\\
 && W(^3\mbox{D}_1) = 0.25 \times 10^{-4} {\mathcal E}^2
\label{Etau}
\end{eqnarray}
For an electric field of 10 kV/cm, the lifetime of the $^3\mbox{D}_2$ 
state is 30 $\mu$s, while the lifetime of the  $^3\mbox{D}_1$ state is
240 ms. This latter result is in good agreement with estimations done by 
Budker and DeMille \cite{Budker}. Note that the state $^3\mbox{D}_2$, with
maximum or minimal possible projection of the total momentum on the 
direction of
the electric field ($M= \pm 2$), cannot be mixed by this field with the
 $^3\mbox{P}_1$ state. Therefore its lifetime is much less affected.

\section{Conclusion}

The radium atom turns out to be a very promising candidate for the study of
time and space invariance violating effects. All such effects considered 
in this paper are strongly enhanced due to the high value of the nuclear
charge $Z$ and the closeness of the opposite parity states of the atom.
Moreover, the contribution of different mechanisms to the time and space 
invariance violating effects can be studied separately if measurements 
are performed for different states and different isotopes of the radium
atom. For example, the atomic EDM induced by the electron EDM is strongly
enhanced in the $^3$D$_1$ state, while contributions of the nuclear Schiff 
and magnetic quadrupole moments are strongly enhanced in the $^3$D$_2$
state. On the other hand, the magnetic quadrupole moment is zero for 
isotopes with nuclear spin $ I = 1/2$, like $^{225}$Ra, while the Schiff
moment for these isotopes is not zero. 

Calculations of the space and time invariance violating effects in radium 
reveal the importance of relativistic and many-body effects. 
The accuracy achieved in the present work is probably 20-30 \%.
However, a further improvement in accuracy is possible if such a need arises
from the progress in measurements.

This work was supported by the Australian Research Council.
%####################################################################
\appendix
%\widetext
\section{Wave functions and matrix elements}
\label{ap}
\subsection{Radium wave functions}

Two-electron wave functions of the ground ($^1\mbox{S}_0$) and three
excited ($^3\mbox{P}_1$,$^3\mbox{D}_1$ and $^3\mbox{D}_2$) states
of radium used in this work for the calculation of space and time
invariance violation have the following form
\begin{eqnarray}
 &&|7s^2 ~~J=0,L=0,M=0 \rangle  = \nonumber \\
 &&-0.9757 |7s_{\frac{1}{2},-\frac{1}{2}}7s_{\frac{1}{2},\frac{1}{2}} \rangle 
  -0.1150 |7p_{\frac{1}{2},-\frac{1}{2}}7p_{\frac{1}{2},\frac{1}{2}}\rangle-
 \nonumber \\
 &&-0.0752(|7p_{\frac{3}{2},-\frac{3}{2}}7p_{\frac{3}{2},\frac{3}{2}} \rangle 
 -       |7p_{\frac{3}{2},-\frac{1}{2}}7p_{\frac{3}{2},\frac{1}{2}}\rangle) +
\label{1S0} \\
 &&+0.0658(|6d_{\frac{3}{2},-\frac{3}{2}}6d_{\frac{3}{2},\frac{3}{2}} \rangle 
 -       |6d_{\frac{3}{2},-\frac{1}{2}}6d_{\frac{3}{2},\frac{1}{2}}\rangle) +
 \nonumber \\
 &&+0.0702(|6d_{\frac{5}{2},-\frac{5}{2}}6d_{\frac{5}{2},\frac{5}{2}} \rangle 
 -       |6d_{\frac{5}{2},-\frac{3}{2}}6d_{\frac{5}{2},\frac{3}{2}}\rangle 
 +       |6d_{\frac{5}{2},-\frac{1}{2}}6d_{\frac{5}{2},\frac{1}{2}}\rangle),
 \nonumber \\
 &&|7s7p ~~J=1,L=1,M=1 \rangle  = \nonumber \\
 &&-0.9010|7s_{\frac{1}{2},\frac{1}{2}}7p_{\frac{1}{2},\frac{1}{2}} \rangle 
  -0.3537|7s_{\frac{1}{2},-\frac{1}{2}}7p_{\frac{3}{2},\frac{3}{2}}\rangle+
   0.2042|7s_{\frac{1}{2},\frac{1}{2}}7p_{\frac{3}{2},\frac{1}{2}} \rangle-
 \nonumber \\
 &&-0.0976|7p_{\frac{1}{2},-\frac{1}{2}}6d_{\frac{3}{2},\frac{3}{2}}\rangle
   +0.0563|7p_{\frac{1}{2}, \frac{1}{2}}6d_{\frac{3}{2},\frac{1}{2}} \rangle 
   +0.0512|7p_{\frac{3}{2},-\frac{1}{2}}6d_{\frac{3}{2},\frac{3}{2}}\rangle-
\label{3P1}  \\
 &&-0.0591|7p_{\frac{3}{2}, \frac{1}{2}}6d_{\frac{3}{2},\frac{1}{2}}\rangle 
  +0.0512|7p_{\frac{3}{2}, \frac{3}{2}}6d_{\frac{3}{2},-\frac{1}{2}}\rangle
  -0.0018|7p_{\frac{3}{2},-\frac{3}{2}}6d_{\frac{5}{2},\frac{5}{2}}\rangle+ 
 \nonumber \\
 && +0.0014|7p_{\frac{3}{2},-\frac{1}{2}}6d_{\frac{5}{2},\frac{3}{2}}\rangle
    -0.0010|7p_{\frac{3}{2}, \frac{1}{2}}6d_{\frac{5}{2},\frac{1}{2}}\rangle 
    +0.0006|7p_{\frac{3}{2}, \frac{3}{2}}6d_{\frac{5}{2},-\frac{1}{2}}\rangle,
 \nonumber \\
 &&|7s6d ~~J=1,L=2,M=1 \rangle  = \nonumber \\
 &&-0.8660|7s_{\frac{1}{2},-\frac{1}{2}}6d_{\frac{3}{2},\frac{3}{2}} \rangle+
    0.5000|7s_{\frac{1}{2}, \frac{1}{2}}6d_{\frac{3}{2},\frac{1}{2}}\rangle+
    0.0002|6d_{\frac{3}{2},-\frac{3}{2}}6d_{\frac{5}{2},\frac{5}{2}} \rangle-
 \label{3D1} \\
 &&-0.0001|6d_{\frac{3}{2},-\frac{1}{2}}6d_{\frac{5}{2},\frac{3}{2}}\rangle
   +0.0001|6d_{\frac{3}{2}, \frac{1}{2}}6d_{\frac{5}{2},\frac{1}{2}} \rangle 
   -0.0001|6d_{\frac{3}{2}, \frac{3}{2}}6d_{\frac{5}{2},-\frac{1}{2}}\rangle-
  \nonumber \\
 &&-0.0021|7p_{\frac{1}{2},-\frac{1}{2}}7p_{\frac{3}{2},\frac{3}{2}}\rangle 
   -0.0012|7p_{\frac{1}{2}, \frac{1}{2}}7p_{\frac{3}{2},\frac{1}{2}}\rangle,
 \nonumber \\
 &&|7s6d ~~J=2,L=2,M=2 \rangle  = \nonumber \\
 &&-0.8087|7s_{\frac{1}{2},\frac{1}{2}}6d_{\frac{3}{2},\frac{3}{2}} \rangle 
   -0.5366|7s_{\frac{1}{2},-\frac{1}{2}}6d_{\frac{5}{2},\frac{5}{2}}\rangle+
    0.2400|7s_{\frac{1}{2},\frac{1}{2}}6d_{\frac{5}{2},\frac{3}{2}} \rangle-
 \nonumber \\
 &&-0.0084|6d_{\frac{3}{2}, \frac{1}{2}}6d_{\frac{3}{2},\frac{3}{2}}\rangle
   -0.0059|6d_{\frac{3}{2},-\frac{1}{2}}6d_{\frac{5}{2},\frac{5}{2}} \rangle 
   +0.0053|6d_{\frac{3}{2}, \frac{1}{2}}6d_{\frac{5}{2},\frac{3}{2}}\rangle-
\label{3D2}  \\
 &&-0.0032|6d_{\frac{3}{2}, \frac{3}{2}}6d_{\frac{5}{2},\frac{1}{2}}\rangle 
   -0.0068|6d_{\frac{5}{2},-\frac{1}{2}}6d_{\frac{5}{2},\frac{5}{2}}\rangle
   +0.0091|6d_{\frac{5}{2}, \frac{1}{2}}6d_{\frac{5}{2},\frac{3}{2}}\rangle+ 
 \nonumber \\
 && +0.0130|7p_{\frac{1}{2}, \frac{1}{2}}7p_{\frac{3}{2},\frac{3}{2}}\rangle
    +0.0038|7p_{\frac{3}{2}, \frac{1}{2}}7p_{\frac{3}{2},\frac{3}{2}}\rangle. 
 \nonumber 
\end{eqnarray}
We use the following form for the single-electron wave function
\begin{eqnarray}
\psi({\bf r})_{jlm} = \frac{1}{r}
 \left( \begin{array}{c}
f(r) \Omega({\bf r}/r)_{jlm} \\ i 
\alpha g(r) \tilde{\Omega}({\bf r}/r)_{jlm}
\end{array}
\right).
\label{psi}
\end{eqnarray}
Here $\alpha = 1/137.036$ is the fine structure constant,
$\tilde{\Omega}({\bf r}/r)_{jlm} = -(\vec \sigma \cdot {\mathbf n}) 
{\Omega}({\bf r}/r)_{jlm}$.

\subsection{Spin-independent weak interaction}

The Hamiltonian of the  spin-independent weak interaction of an electron 
with the nucleus is given by \cite{Khriplovich}
\begin{equation}
	H_{PNC} = - \frac{G}{2\sqrt{2}} \rho (r) Q_W \gamma_5,
\label{APNC}
\end{equation}
where $G=2.22255 \times 10^{-14}$ a.u. is the Fermi constant, $\rho$ is the 
nuclear density ($\int \rho dV =1$), $Q_W \approx -N + Z(1-4 \sin^2\theta_W)$ 
is the nuclear weak charge, and $\gamma_5$ is a Pauli matrix.
The matrix element of (\ref{APNC}) with wave functions (\ref{psi}) has a form
\begin{eqnarray}
	& \langle j_1 l_1 m_1 |H_{PNC}| j_2 l_2 m_2 \rangle =
	-i \frac{G}{2\sqrt{2}} Q_W R_{PNC}
	\delta_{j_1 j_2} \delta_{l_1 \tilde l_2} \delta_{m_1 m_2},\\
\label{apnc}
	& R_{PNC} =  \alpha \int \rho (f_1 g_2 - g_1 f_2) dr - 
	\mbox{radial integral} \nonumber, \\
	& \tilde l = 2j -l. \nonumber
\end{eqnarray}
However, it is often more convenient to express (\ref{apnc}) in a form 
\begin{eqnarray}
	\langle j_1 l_1 m_1 |H_{PNC}| j_2 l_2 m_2 \rangle =
	(-1)^{j_1-m_1} \left( \begin{array}{ccc} j_1 & 0 & j_2 \\
	-m_1 & 0 & m_2 \\ \end{array} \right)
	(-i) \alpha \frac{G}{2\sqrt{2}} Q_W C_{PNC} R_{PNC}, \\
	C_{PNC} = \sqrt{2j_1+1} \delta_{j_1 j_2} \delta_{l_1 \tilde l_2}
	- \mbox{angular coefficient for the reduced matrix element.}
	\nonumber
\end{eqnarray}

%--------------------------------------------------------------------------
\subsection{Anapole moment}

The Hamiltonian of the interaction of an electron with the nuclear 
anapole moment has the form \cite{Flambaum80}
\begin{equation}
	H_{AM} = \frac{G}{\sqrt{2}} \frac{(\mathbf{I}\cdot \vec \alpha)}
	{I(I+1)} K \kappa_a \rho (r),
\label{AAM}
\end{equation}
where  $I$ is the nuclear spin,
$K=(I+\frac{1}{2})(-1)^{I+\frac{1}{2}-l}, l$ is the orbital momentum of the
outermost nucleon, $\kappa_a$ is a dimensionless constant proportional to
the strength of the nucleon-nucleon PNC interaction \cite{FKS}. 
The matrix 
elements of the Hamiltonian (\ref{AAM}) between the many-electron states
of the atoms depend on the hyperfine structure (see, e.g. \cite{Varshalovich})
\begin{eqnarray}
  & \langle IJ'F |H_{AM}| IJF \rangle =
\frac{G}{\sqrt{2}} \frac{K \kappa_a}{I(I+1)}(-1)^{F+I+J'}
\left\{ \begin{array}{ccc} I & I & 1 \\ J & J' & F \end{array} \right\}
\times \nonumber \\
 & \times \sqrt{I(I+1)(2I+1)}\langle J' ||\vec \alpha \rho(r)|| J \rangle, \\
 & {\mathbf F = I + J, J} - \mbox{atomic total momentum} \nonumber.
\end{eqnarray}
The electron part of the operator (\ref{AAM}) is $\vec \alpha \rho(r)$.
Its single-electron matrix elements over states (\ref{psi}) have a form
\begin{eqnarray}
	& \langle j_1 l_1 m_1 |\vec \alpha \rho(r)| j_2 l_2 m_2 \rangle =
	(-1)^{j_1-m_1} \left( \begin{array}{ccc} j_1 & 1 & j_2 \\
	-m_1 & q & m_2 \\ \end{array} \right)
	(C_{1AM}R_{1AM} + C_{2AM}R_{2AM}), \\
	& C_{1AM} = (-1)^{j_1+l_2+\frac{1}{2}}\sqrt{6(2j_1+1)(2j_2+1)}
	\left\{ \begin{array}{ccc} \frac{1}{2} & j_1 & l_2 \\
	j_2 & \frac{1}{2} & 1 \end{array} \right\} \delta_{\tilde l_1l_2}
	, \nonumber \\
	& C_{2AM} = (-1)^{j_1+l_1+\frac{3}{2}}\sqrt{6(2j_1+1)(2j_2+1)}
	\left\{ \begin{array}{ccc} \frac{1}{2} & j_1 & l_1 \\
	j_2 & \frac{1}{2} & 1 \end{array} \right\} \delta_{l_1\tilde l_2}
	, \nonumber \\
	& R_{1AM} = - 4\pi \alpha \int g_1 f_2 dr, \nonumber \\
	& R_{2AM} = - 4\pi \alpha \int f_1 g_2 dr. \nonumber 
\end{eqnarray}

The dominating contribution to the $z$-component of the parity non-conserving 
electric dipole transition amplitude between the $^1$S$_0$ and $^3$D$_1$ 
states of Ra induced by the anapole moment is given by
\begin{eqnarray}
& E1_{PV} = (-1)^{F-f} \left( \begin{array}{ccc} F & 1 & F' \\
 -f & 0 & f \\ \end{array} \right) (-1)^{4F'+J+J'+2I+1}
\frac{G}{\sqrt{2}}K\kappa_a \sqrt{\frac{2I+1}{I(I+1)}} \times \nonumber \\
& \times \sqrt{(2F+1)(2F'+1)}\left\{\begin{array}{ccc} J' & I & F' \\
F & 1 & J \\ \end{array} \right\} \left\{ \begin{array}{ccc}
I & I & 1 \\ J & J' & F' \\ \end{array} \right\}
\frac{\langle 7s^2 ||E1||7s7p \rangle \langle 7s7p ||\vec \alpha \rho(r)
||7s6d \rangle}{E_{7s6d} - E_{7s7p}}.
\label{AE1AM}
\end{eqnarray}
Here ${\mathbf F=I+J}, f=\min(F,F')$.

%------------------------------------------------------------------
\subsection{Electron EDM}

The Hamiltonian of the interaction of the {\bf electron EDM} $d_e$
with the atomic electric field ${\mathbf E}$ has the form \cite{Khriplovich}
\begin{eqnarray}
	& H_{EDM} = -d_e \beta ({\mathbf \Sigma \cdot E}),
\label{AEDM1}
\end{eqnarray}
where
\[
	\beta = \left( \begin{array}{cc} 1 & 0 \\
	0 & -1 \end{array} \right), ~~{\mathbf \Sigma} = 
	\left( \begin{array}{cc}
	\vec \sigma & 0 \\ 0 & \vec \sigma \end{array} \right), 
	~~ {\mathbf E = - \nabla} V({\mathbf r}). 
\]
The atomic EDM induced by (\ref{AEDM1}) can be calculated as an average
value of the operator of the dipole moment over states mixed by an operator
similar to (\ref{AEDM1})
\begin{eqnarray}
	& H'_{EDM} = -d_e (\beta-1) ({\mathbf \Sigma \cdot E}).
\label{AEDM2}
\end{eqnarray}
Its single-electron matrix elements have a form
\begin{eqnarray}
	& \langle j_1 l_1 m_1 |H'_{EDM}| j_2 l_2 m_2 \rangle =
	(-1)^{j_1-m_1} \left( \begin{array}{ccc} j_1 & 0 & j_2 \\
	-m_1 & 0 & m_2 \\ \end{array} \right)
	d_e C_{EDM} R_{EDM}, \\
	& C_{EDM} = \sqrt{2j_1+1} \delta_{j_1 j_2} \delta_{l_1 \tilde l_2},
	\nonumber \\
	& R_{EDM} =  2\alpha^2 \int g_1 \frac{dV}{dr} g_2 dr. 
	\nonumber
\end{eqnarray}
Note that the selection rules and the angular coefficients are the same as 
for the spin independent weak interaction (\ref{APNC}), while the radial 
integrals are different.

%------------------------------------------------------------------------
\subsection{Schiff moment}

The Hamiltonian of the interaction of an electron with the nuclear 
Schiff moment has the form \cite{Sushkov84}
\begin{equation}
	H_{SM} = 4\pi {\mathbf S \cdot \nabla} \rho (r),
\label{ASM}
\end{equation}
${\mathbf S} = S {\mathbf I}/I, S$ is Schiff moment.
Many-electron matrix elements of (\ref{ASM}) depend on the hyperfine 
structure similar to (\ref{AAM})
\begin{eqnarray}
 & \langle IJ'F |H_{SM}| IJF \rangle =
(-1)^{F+I+J'}
\left\{ \begin{array}{ccc} I & I & 1 \\ J & J' & F \end{array} \right\}
S \sqrt{\frac{I(I+1)(2I+1)}{I}}
\langle J' ||4\pi {\mathbf \nabla} \rho(r)|| J \rangle.
\label{SMF}
\end{eqnarray}
The electron part of the operator (\ref{ASM}) is $4\pi {\mathbf \nabla} 
\rho(r)$.
Its single-electron matrix elements over states (\ref{psi}) have the form
\begin{eqnarray}
	& \langle j_1 l_1 m_1 |4\pi {\mathbf \nabla} \rho(r)| j_2 l_2 m_2 
	\rangle =
	(-1)^{j_1-m_1} \left( \begin{array}{ccc} j_1 & 1 & j_2 \\
	-m_1 & q & m_2 \\ \end{array} \right)
	C_{SM}R_{SM}, \\
\label{WSM}
	& C_{SM} = (-1)^{j_2 + \frac{3}{2}} \sqrt{(2j_1+1)(2j_2+1)}
	\left( \begin{array}{ccc} j_1 & j_2 & 1 \\ 
	\frac{1}{2} & \frac{1}{2} & 0 \\ \end{array} \right) \xi(l_1+l_2+1),
	\nonumber \\
	& \xi(x)= \left\{ \begin{array}{ll} 1, & \mbox{if}~x~\mbox{is even} \\
	 0, & \mbox{if}~x~\mbox{is odd} \end{array} \right. , \nonumber \\
	& R_{SM} = -4\pi \int (f_1f_2+\alpha^2g_1g_2)\frac{d\rho}{dr} dr.
	\nonumber
\end{eqnarray}

The EDM of Ra induced by the nuclear Schiff moment for a particular hyperfine 
structure component of the $^3$D$_2$ state is approximately given by
\begin{eqnarray}
 & d_z = 2 \left( \begin{array}{ccc} F & 1 & F \\
 -F & 0 & F \\ \end{array} \right) (-1)^{2F+2I+J+J'}
\left\{\begin{array}{ccc} J' & I & F \\
F & 1 & J \\ \end{array} \right\} \left\{ \begin{array}{ccc}
I & I & 1 \\ J & J' & F \\ \end{array} \right\}\sqrt{\frac{(I+1)(2I+1)}{I}}
 \times \nonumber \\
	& (2F+1)S
\frac{\langle 7s6d~^3\mbox{D}_J ||E1||7s7p~^3\mbox{P}_{J'} \rangle 
\langle 7s7p~^3\mbox{P}_{J'}||4\pi{\mathbf \nabla}\rho(r)||7s6d~^3\mbox{D}_J 
\rangle}
{E_{7s6d} - E_{7s7p}}.
\label{AEDMSM}
\end{eqnarray}
%-------------------------------------------------------------------------
\subsection{Magnetic quadrupole moment}

The Hamiltonian of the interaction of an electron with the nuclear 
magnetic quadrupole moment has the form \cite{Sushkov84}
\begin{eqnarray}
	H_{MQM}& = &- \frac{M}{4I(2I-1)} t_{mk} A_{mk}, \nonumber \\
	t_{mk}& = & I_m I_k + I_k I_m - \frac{2}{3} \delta_{km} I(I+1), \\
\label{HMQM}
	A_{mk}& = &\epsilon_{nim} \alpha_n \partial_i \partial_k 
	\frac{1}{r}.
	\nonumber 
\end{eqnarray}
Its many-electron matrix element is
\begin{eqnarray}
	 & \langle IJ'F |H_{MQM}| IJF \rangle = \frac{3M}{8I(2I-1)}
	\sqrt{5(2F+1)(2I+3)(I+1)(2I+1)I(2I-1)} \times \nonumber \\
	& \left\{ \begin{array}{ccc} 2 & 2 & 0 \\ J' & I & F \\
	J & I & F \end{array} \right\} \langle J'||A_{mk}||J \rangle. \\
\label{MQMF}
\end{eqnarray}
The single-electron matrix elements of the operator $A_{mk}$ have the form
\begin{eqnarray}
	& \langle j_1 l_1 m_1 |A_{mk}| j_2 l_2 m_2 
	\rangle =
	(-1)^{j_1-m_1} \left( \begin{array}{ccc} j_1 & 2 & j_2 \\
	-m_1 & q & m_2 \\ \end{array} \right)
	(C_{1MQM}+C_{2MQM})R_{MQM}, \\
\label{WMQM}
	& C_{1MQM} = (-1)^{j_2 - \frac{1}{2}} 
	\frac{4}{3}\sqrt{(2j_1+1)(2j_2+1)}
	\left( \begin{array}{ccc} j_1 & j_2 & 2 \\ 
	\frac{1}{2} & \frac{1}{2} & 0 \\ \end{array} \right) \xi(l_1+l_2+1),
	\nonumber \\
	& C_{2MQM} = (-1)^{j_1+j_2+l_2+1} 
	4\sqrt{5(2j_1+1)(2j_2+1)(2l_1+1)(2l_2+1)} \times \nonumber \\
	& \left( \begin{array}{ccc} l_1 & 1 & l_2 \\ 0 & 0 & 0 \end{array}
	\right) \left\{ \begin{array}{ccc} 1 & l_1 & l_2 \\
	2 & j_1 & j_2 \\ 1 & \frac{1}{2} & \frac{1}{2} \end{array} \right\},
	\nonumber \\
	& R_{MQM} = \alpha \int F(r)(g_1 f_2 + f_1 g_2)dr, \nonumber \\
	& \mbox{where} F(r) = \left\{ \begin{array}{ll} r/r_N^4, & 
	\mbox{if}~r \le r_N \\ 1/r^3, & \mbox{if}~r > r_N 
	\end{array} \right. , \nonumber \\
	& r_N - \mbox{nuclear radius}. \nonumber
\end{eqnarray}
The EDM of Ra induced by the nuclear MQM for a particular hyperfine 
structure component of the $^3$D$_J$ state is approximately given by
\begin{eqnarray}
d_z & = & 2 \left( \begin{array}{ccc} F & 1 & F \\
 -F & 0 & F \\ \end{array} \right) (-1)^{F+I+J}(2F+1)^{\frac{3}{2}}
\frac{3\sqrt{5}}{4}
M\sqrt{\frac{(2I+3)(I+1)(2I+1)}{I(2I-1)}} \times \nonumber \\
 &&\left\{\begin{array}{ccc} J' & I & F \\
F & 1 & J \\ \end{array} \right\} \left\{ \begin{array}{ccc}
2 & 2 & 0 \\ J & I & F \\ J' & I & F \\ \end{array} \right\}
\frac{\langle 7s6d~^3\mbox{D}_J ||E1||7s7p~^3\mbox{P}_{J'} \rangle 
\langle 7s7p~^3\mbox{P}_{J'}||A_{mk}||7s6d~^3\mbox{D}_J 
\rangle}
{E_{7s6d} - E_{7s7p}}.
\label{AEDMMQM}
\end{eqnarray}
For the case of the EDM in the $^3$D$_J$ state, $J'=1, J=2$ in (\ref{AEDMMQM}).
%**************************************************************************

%**********************************************************************   
\begin{table}
\caption{Energies and hyperfine structure constants of lower excited states of 
$^{137}$Ba ($I$=3/2 $\mu$=0.937365) and $^{213}$Ra ($I$=1/2,$\mu$=0.6133).}
\label{BaRa}
\begin{tabular}{crrrrd}
Atom & State & \multicolumn{2}{c}{Energies (cm$^{-1}$)} & 
\multicolumn{2}{c}{hfs constant A (MHz)} \\
     &       & Calc. & Exper.\cite{Moore} & Calc. & Exper. \cite{hfs} \\
\hline
 Ba & $6s5d~~ ^3$D$_1$ &  9225 &  9034 & -632  & -520.5  \\
    & $       ^3$D$_2$ &  9346 &  9216 &  357  &  415.9  \\
    & $       ^3$D$_3$ &  9554 &  9596 &  504  &  456.6  \\
    & $       ^1$D$_2$ & 12147 & 11395 &  -26  & -82.18  \\
    & $6s6p~~ ^3$P$_0$ & 12203 & 12226 &       &         \\
    & $       ^3$P$_1$ & 12577 & 12637 &  1233 & 1150.59 \\
    & $       ^3$P$_2$ & 12464 & 13514 &   878 &    \\
    & $       ^1$P$_1$ & 18042 & 18060 &   -48 & -109.2   \\
 Ra & $7s7p~~ ^3$P$_0$ & 12971 & 13078 &       &    \\
    & $       ^3$P$_1$ & 13926 & 13999 &  8058 &    \\
    & $       ^3$P$_2$ & 16660 & 16689 &  4637 &    \\
    & $       ^1$P$_1$ & 21033 & 20716 & -1648 & -2315 \\
    & $7s6d~~ ^3$D$_1$ & 13893 & 13716 & -4108 &    \\
    & $       ^3$D$_2$ & 14042 & 13994 &  1749 &    \\
    & $       ^3$D$_3$ & 14299 & 14707 &  2744 &    \\
    & $       ^1$D$_2$ & 17750 & 17081 &  -320 &    \\
\end{tabular}
\end{table}
%*******************************************************************
\begin{table}
\caption{Single-electron matrix elements of the $P$- and $T$-odd
interactions for radium (presented reduced matrix elements of an
electron part of the Hamiltonian as specified in the table, see Appendix
for details). All values are in atomic units.}
\label{rad}
\begin{tabular}{cccc}
Matrix element & \multicolumn{2}{c}{Approximation} & Even or\\
               & RHF\tablenotemark[1] &  RHF+RPA+$\Sigma$\tablenotemark[2]
 & odd\tablenotemark[3] \\
\hline
\multicolumn{4}{c}{Spin-independent PNC interaction, $H=\rho(r)\gamma_5$} \\
$\langle 7s_{1/2}||H||7p_{1/2} \rangle$ & -2769 & -3832 & Odd \\
$\langle 7p_{3/2}||H||6d_{3/2} \rangle$ & 0.004 & -146.8 & Odd \\
\multicolumn{4}{c}{Nuclear Anapole moment, $H=\vec \alpha \rho(r)$} \\
$\langle 7s_{1/2}||H||7p_{1/2} \rangle$ & -503 & -577 & Odd \\
$\langle 7s_{1/2}||H||7p_{3/2} \rangle$ & -0.508 & 20.26 & Even \\
$\langle 7p_{1/2}||H||6d_{3/2} \rangle$ & -0.024 & -66.29 & Even \\
$\langle 7p_{3/2}||H||6d_{3/2} \rangle$ & 0.0006 & -29.99 & Odd \\
$\langle 7p_{3/2}||H||6d_{5/2} \rangle$ & 0      & 11.71 & Even \\
\multicolumn{4}{c}{Electron dipole moment, $H=(\beta-1){\mathbf \Sigma E}$} \\
$\langle 7s_{1/2}||H||7p_{1/2} \rangle$ & 12.06 & 17.05 & Even \\
$\langle 7p_{3/2}||H||6d_{3/2} \rangle$ & 0.556 & 2.082 & Even \\
\multicolumn{4}{c}{Nuclear Schiff moment, $H=4\pi{\mathbf \nabla}\rho(r)$} \\
$\langle 7s_{1/2}||H||7p_{1/2} \rangle$ & -44400 & -63027 & Even \\
$\langle 7s_{1/2}||H||7p_{3/2} \rangle$ & -32550 & -56730 & Odd \\
$\langle 7p_{1/2}||H||6d_{3/2} \rangle$ & -1497  & 1873   & Odd \\
$\langle 7p_{3/2}||H||6d_{3/2} \rangle$ & -0.03  &  2767  & Even \\
$\langle 7p_{3/2}||H||6d_{5/2} \rangle$ & -0.07  &  8163 & Even \\
\multicolumn{4}{c}{Nuclear Magnetic quadrupole moment, $H=A_{mk}$} \\
$\langle 7s_{1/2}||H||7p_{3/2} \rangle$ & 17.28 & 25.06 & Odd \\
$\langle 7p_{1/2}||H||6d_{3/2} \rangle$ & 2.831 & 2.933 & Odd \\
$\langle 7p_{1/2}||H||6d_{5/2} \rangle$ & -0.2017 & 6.631 & Even \\
$\langle 7p_{3/2}||H||6d_{5/2} \rangle$ & 0.5389 & 4.011 & Odd \\
\end{tabular}
\tablenotetext[1]{Relativistic Hartree-Fock}
\tablenotetext[2]{Core polarization and core-valence correlation interaction 
are included}
\tablenotetext[3]{Even means that 
$\langle i||H||j \rangle = \langle j||H||i \rangle$; odd means that
$\langle i||H||j \rangle = - \langle j||H||i \rangle$.}
\end{table}
%****************************************************************
\begin{table}
\caption{E1-transition amplitudes for Ba and Ra 
($|\langle i||d_z|| j \rangle| a_0$).}
\label{E1}
\begin{tabular}{ccdddd}
\multicolumn{2}{c}{Transition} & \multicolumn{2}{c}{Ba} & 
\multicolumn{2}{c}{Ra}\\
   $i$ & $j$  & Amplitude &   \multicolumn{1}{c}{Frequency $\epsilon_i
 - \epsilon_j$ (a.u.)} & 
Amplitude & \multicolumn{1}{c}{Frequency  $\epsilon_i - \epsilon_j$ (a.u.)} \\
\hline
 $^3$P$_0$ & $^3$D$_1$ & 2.3121 & 0.01473 & 3.0449 & -0.002904 \\
 $^3$P$_1$ & $^1$S$_0$ & 0.4537 & 0.05758 & 1.0337 &  0.06379 \\
 $^3$P$_1$ & $^3$D$_1$ & 2.0108 & 0.01641 & 2.6389 & 0.001292  \\
 $^3$P$_1$ & $^3$D$_2$ & 3.4425 & 0.01559 & 4.4399 & 0.0000247 \\
 $^3$P$_1$ & $^1$D$_2$ & 0.1610 & 0.00566 & 0.0467 & -0.01404 \\
 $^3$P$_2$ & $^3$D$_1$ & 0.5275 & 0.02042 & 0.7166 &  0.01354 \\
 $^3$P$_2$ & $^3$D$_2$ & 2.024  & 0.01959 & 2.7283 &  0.01228 \\
 $^3$P$_2$ & $^3$D$_3$ & 4.777  & 0.01785 & 6.3728 &  0.009027 \\
 $^3$P$_2$ & $^1$D$_2$ & 0.1573 & 0.00966 & 0.1499 & -0.001790 \\
 $^1$P$_1$ & $^1$S$_0$ & 5.236  & 0.08229 & 5.4797 & 0.09439 \\
 $^1$P$_1$ & $^3$D$_1$ & 0.1047 & 0.04113 & 0.4441 & 0.03189 \\
 $^1$P$_1$ & $^3$D$_2$ & 0.4827 & 0.04030 & 1.188  & 0.03063 \\
 $^1$P$_1$ & $^1$D$_2$ & 1.047  & 0.03037 & 2.4053 & 0.01656 \\
\end{tabular}
\end{table}
%****************************************************************
\begin{table}
\caption{Single-electron contributions to the two-electron matrix
element $\langle ^3\mbox{P}_1 ||H||^3\mbox{D}_2 \rangle$.
Dash means no contribution due to selection rules. Zero means very small
contribution. Same units as in Table \ref{rad}.} 
\label{HME}
\begin{tabular}{ccccc}
Transition & $H=-e{\mathbf r}\tablenotemark[1] $ & 
$H={\mathbf \alpha}\rho(r)$\tablenotemark[2] & 
$H= 4\pi\nabla\rho(r)$\tablenotemark[3] &
$H= A_{mk}$\tablenotemark[4] \\
\hline
$7s_{1/2} - 7p_{1/2}$ & -0.3677 &  58.78 &  6421 &  - \\
$7p_{1/2} - 7s_{1/2}$ &  0.0215 &  3.431 &  -375 &  - \\
$7s_{1/2} - 7p_{3/2}$ & -0.1306 & -0.515 &  1441 & -1.565 \\
$7p_{3/2} - 7s_{1/2}$ & -0.0585 &  0.230 &   645 & -0.289 \\
$7p_{1/2} - 6d_{3/2}$ &  0.0020 &  0.029 &    -1 & -0.003 \\
$6d_{3/2} - 7p_{1/2}$ &  3.856  & -54.08 & -1528 & -1.848 \\
$7p_{1/2} - 6d_{5/2}$ &    -    &    -   &    -  &  0     \\
$6d_{5/2} - 7p_{1/2}$ &    -    &    -   &    -  & -2.032 \\
$7p_{3/2} - 6d_{3/2}$ &  0.0004 &  0.006 &    -1 &   -    \\
$6d_{3/2} - 7p_{3/2}$ & -0.2470 &  3.522 &   325 &   -    \\
$7p_{3/2} - 6d_{5/2}$ &  0      &  0     &     0 &    0   \\
$6d_{5/2} - 7p_{3/2}$ &  1.364  &  2.442 & -1702 & -0.736 \\
\hline
 Total                &  4.4399 &  13.85 &  5226 & -6.473 \\
%                          E1       AnM     SM     MQM
\end{tabular}
\tablenotetext[1]{For $E1$ transition amplitude}
\tablenotetext[2]{For anapole moment contribution}
\tablenotetext[3]{For Schiff moment contribution}
\tablenotetext[4]{For Magnetic quadrupole moment contribution}
\end{table}
%****************************************************************
\begin{table}
\caption{Parity non-conserving E1-transition amplitude induced by
nuclear anapole moment}
\label{AnM}
\begin{tabular}{ccccc}
 $I$ & $F$ & $F'$ & \multicolumn{2}{c}{
$\langle d_z \rangle$ in units $10^{-10}\kappa_a iea_0$} \\
   &   &    &   $^1$S$_0 - ^3$D$_1$ &  $^1$S$_0 - ^3$D$_2$ \\
\hline
 0.5 & 0.5 & 1.5 & 2.05 & -20.3 \\
 1.5 & 1.5 & 0.5 & -0.58 & 5.7 \\
     & 1.5 & 1.5 & -1.4 & 13.8 \\
     & 1.5 & 2.5 &  1.3 & -12.9 \\
\end{tabular}
\end{table}
%****************************************************************
\begin{table}
\caption{EDM of Ra atom in the $^3$D$_2$ state induced by nuclear 
Schiff moment}
\label{RSM}
\begin{tabular}{ccccc}
 $I$ & $F$ & \multicolumn{2}{c}{$d_z$ (a.u.)} & $d_z (e cm)$ \\
\hline
 0.5 & 1.5 & $-0.94\times10^8S$ & $-0.19\times10^{-11}\eta$\tablenotemark[1] 
 & $-0.36\times10^{-19}\eta$ \\
 1.5 & 0.5 & $-0.16\times10^8S$ & $-0.42\times10^{-11}\eta$\tablenotemark[2] 
 & $-0.80\times10^{-19}\eta$ \\
 1.5 & 1.5 & $-0.30\times10^9S$ & $-0.81\times10^{-11}\eta$\tablenotemark[2] 
 & $-0.15\times10^{-18}\eta$ \\
 1.5 & 2.5 & $-0.28\times10^9S$ & $-0.76\times10^{-11}\eta$\tablenotemark[2] 
 & $-0.14\times10^{-18}\eta$ \\
\end{tabular}
\tablenotetext[1]{Nuclear Schiff moment $S$ is assumed to be 
$S=400\times 10^8 \eta~e~\mbox{fm}^3$ \cite{Spevak}}
\tablenotetext[2]{$S=300\times 10^8 \eta~e~\mbox{fm}^3$ \cite{Spevak}}
\end{table}
%****************************************************************
\begin{table}
\caption{EDM of $^{223}$Ra isotope ($I = 3/2$) in the $^3$D$_2$ state 
induced by nuclear magnetic quadrupole moment}
\label{RMQM}
\begin{tabular}{cccc}
 $F$ & $d_z$\tablenotemark[1] & $d_z$\tablenotemark[2] \\
\hline
 0.5 &  $1344M m_e$ & $7.4 \times 10^{-20} \eta~e \cdot$ cm \\
 1.5 &  $1292M m_e$ & $7.0 \times 10^{-20} \eta~e \cdot$ cm \\
 2.5 &  $-806M m_e$ & $-4.4 \times 10^{-20} \eta~e \cdot$ cm \\
\end{tabular}
\tablenotetext[1]{In terms of nuclear magnetic quadrupole moment $M$}
\tablenotetext[2]{$M$ is assumed to
be $M=10^{-19} (\eta/m_p)~e \cdot$ cm, \cite{Flambaum94} where $m_p$ is the
proton mass}
\end{table}
%*******************************************************************
\begin{table}
\caption{Lifetimes of lower short-living states of Ba and Ra}
\label{tau}
\begin{tabular}{cclcc}
Atom & State & Lower states to decay to  & 
\multicolumn{2}{c}{Lifetime} \\
     &       &  via E1-transitions   & This work & Other data \\
\hline
Ba & $^3$P$_0$ & $^3$D$_1$                & 2.83 $\mu$s & \\ 
   & $^3$P$_1$ & $^1$S$_0,^3$D$_1,^3$D$_2,^1$D$_2$ &
 1.37  $\mu$s & 1.2 $\mu$s \tablenotemark[1] \\ 
   & $^3$P$_2$ & $^3$D$_1,^3$D$_2,^3$D$_3,^1$D$_2$ &
 1.41 $\mu$s & \\ 
   & $^1$P$_1$ & $^1$S$_0,^3$D$_1,^3$D$_2,^1$D$_2$ &
 9.1 ns & 8.5 ns \tablenotemark[1]\\
Ra & $^3$P$_1$ & $^1$S$_0,^3$D$_1,^3$D$_2$ & 505 ns & 
420 ns \tablenotemark[2], 250 ns \tablenotemark[3]  \\ 
   & $^3$P$_2$ & $^3$D$_1,^3$D$_2,^3$D$_3$ &  74.6 ns & \\
   & $^3$D$_1$ & $^3$P$_0$ &  617 $\mu$s & 800 $\mu$s \tablenotemark[4] \\ 
   & $^1$D$_2$ & $^3$P$_1,^3$P$_2$ &  38 ms &  \\ 
   & $^1$P$_1$ & $^1$S$_0,^3$D$_1,^3$D$_2,^1$D$_2$ &
 5.5 ns & \\ 
\end{tabular}
\tablenotetext[1]{Reference \cite{Radzig}}
\tablenotetext[2]{Reference \cite{Hafner}}
\tablenotetext[3]{Reference \cite{Bruneau}}
\tablenotetext[4]{Estimation, Reference \cite{Budker}}
\end{table}
%####################################################################
\end{document}